\documentclass[dvips]{article}

\usepackage{icrc2011}

\title{BL Lac Objects: Laboratories to study the environment and properties of emitting particles in relativistic jets}

\newcommand{\etal}{\MakeLowercase{\textit{et al. }Emitting particles in Blazar jets}} 
\shorttitle{Mankzuhiyil \etal }

\authors{Nijil Mankuzhiyil$^{1}$, Stefano Ansoldi$^{2}$, Massimo Persic$^{3}$, Fabrizio Tavecchio$^{4}$  }
\afiliations{%
$^1$University of Udine \& INFN Trieste, Italy\\
$^2$ ICRA, Roma \& INFN Trieste \& University of Udine, Italy\\
$^3$INAF \& INFN Trieste, Italy\\
$^4$INAF Milano, Italy}
\email{nijil.mankuzhiyil@uniud.it}

\abstract{

We report the variation of the spectral energy distribution (SED) of blazars as a function of source activity, based on available, simultaneous multi-wavelength (MWL) observations of BL Lac objects. We use a fully automatized $\chi ^{2}$ minimization procedure, instead of the commonly used eye-ball fit, to model the data sets with a one-zone Synchrotron-Self-Compton (SSC) model. The obtained SSC parameters are then analyzed as a function of source luminosity, and the correlation between parameters is shown. Possibilities of improving the present observational and modeling status of BL Lac objects are also discussed.}

\begin{document}
\maketitle

\section{Introduction}


Active Galactic Nuclei (AGNs) are believed to be the most powerful energy sources in the universe. It is usually assumed that at the AGN center there is a super massive black hole (SMBH) surrounded by the accretion disk: a beamed emission perpendicular to the disk is expelled in opposite directions in $\sim$10\% of the cases (radio-loud AGNs). The energy required for the beamed emission might be coming from the rotating black hole and the fully ionized accretion disk \cite{ref1}. AGNs are phenomenologically classified according to their viewing angle to the observer. If the jet is pointing very close to the line of sight to the observer the source is called a blazar. Blazars include Flat Spectrum Radio Quasars (FSRQ) and BL Lac objects.

Blazar emission is dominated by a non-thermal continuum. The spectral energy distribution (SED) shows two peaks. For the subclass of high-peaked BL-Lac objects (HBLs) the lower-energy peak is in the UV/X-ray range and the higher-energy peak is in multi-GeV bands; the so called  low-peaked BL Lac objects (LBLs) have, instead, the low-energy peak in the IR/optical band and the high-energy peak in the multi-MeV band. The low-energy peak is due to synchrotron radiation. According to the leptonic model, the high-energy peak is produced by inverse Compton (IC) upscattering of lower-energy photons off relativistic electrons: in particular the so called synchrotron self Compton (SSC) model suggests that the synchrotron photons are upscattered by the same electron population responsible for the synchrotron emission \cite{ref2}.

\section{$\chi^2$-minimized one-zone SSC model}

In our analysis we will employ a specific one-zone SSC model \cite{ref3}. This model assumes that the source is a spherical blob of plasma of radius $R$, moving with a Doppler factor $\delta$ towards the observer at an angle $\theta$  with respect to the line of sight, threaded with a uniform tangled magnetic field of strength $B$. The injected relativistic particle population is described as a broken power-law spectrum with normalization $K$, extending from $\gamma_{\rm min}$ to $\gamma_{\rm max}$ with indices  $n_{\rm 1}$ and $n_{\rm 2}$ below and above the break Lorentz factor $\gamma_{\rm break}$. So, the total number of free parameter of the model is 9: 6 parameters specify the properties, and 3 the environment, of the emitting electrons. The one-zone SSC model can be fully constrained by using simultaneous multi-frequency broad-band data.

In the present work we have kept $\gamma _{\mathrm{min}}$ fixed and equal to $1$, as usually done in the literature. The determination of the remaining parameters has been performed by finding their best values and uncertainties from a $\chi ^{2}$-minimization in which multi-frequency experimental points have been fitted to the SSC model described in \cite{ref3}. Minimization has been performed using the Levenberg-Marquardt method \cite{ref4}, which is an efficient standard for non-linear least-squares minimization that smoothly interpolates between two different minimization approaches, namely the inverse Hessian method and the steepest descent method.

A crucial point in our implementation is that from \cite{ref3} we can only obtain a numerical approximation to the SSC spectrum, in the form of a
sampled SED. On the other hand, at each step of the loop the calculation of $\chi ^{2}$ requires the evaluation of the SED for all the observed frequencies. Although an observed point will likely not be one of the sampled points coming from \cite{ref3}, it will fall between two sampled points, so that interpolation can be used to approximate the value of the SED. At the same time, the Levenberg-Marquardt method requires the calculation of the partial derivatives of $\chi ^{2}$ with respect to the SSC parameters. These derivatives have also been obtained numerically by evaluating the incremental ratio of the $\chi ^{2}$ with respect to a sufficiently small, dynamically adjusted increment of each parameter.

\section{Data selection}

In order to study the behavior of SSC parameters for each source in different emission states, we selected sources with at least 2 available simultaneous SED data and with known redshift. The MWL campaign period of each data sets used for this study is given in Table 1.

\begin{table}[t]
\begin{center}
\begin{tabular}{l|ccc}
\hline
Blazar & State &  Year of Campaign       &  Reference \\
\hline
Mkn\,501 & 1 & 2006  & \cite{ref5}    \\
&2 & 2009 & \cite{ref6}\\
PKS\,2155-304  & 1 & 2006   & \cite{ref7} \\
&2 & 2006 & \cite{ref7} \\
&3 & 2003 & \cite{ref7} \\
&4 &2008 & \cite{ref8} \\
W Comae  & 1 & 2008  & \cite{ref9}     \\
&2 & 2008 &\cite{ref10}\\
1ES\,1959+650  & 1 & 2006  & \cite{ref11}     \\
&2 & 2002 & \cite{ref12}\\
1ES\,1101-232  & 1 & 2005  & \cite{ref13}     \\
&2 & 2004 & \cite{ref13}\\
Mkn\,421 & 1-9 & multiple years & \cite{ref14} \\
\hline
\end{tabular}
\caption{\label{table_single}Log of SED data sets.}
\end{center}
\end{table}
\begin{figure*}[th]
  \centering
  \includegraphics[width=5in,height=4.in]{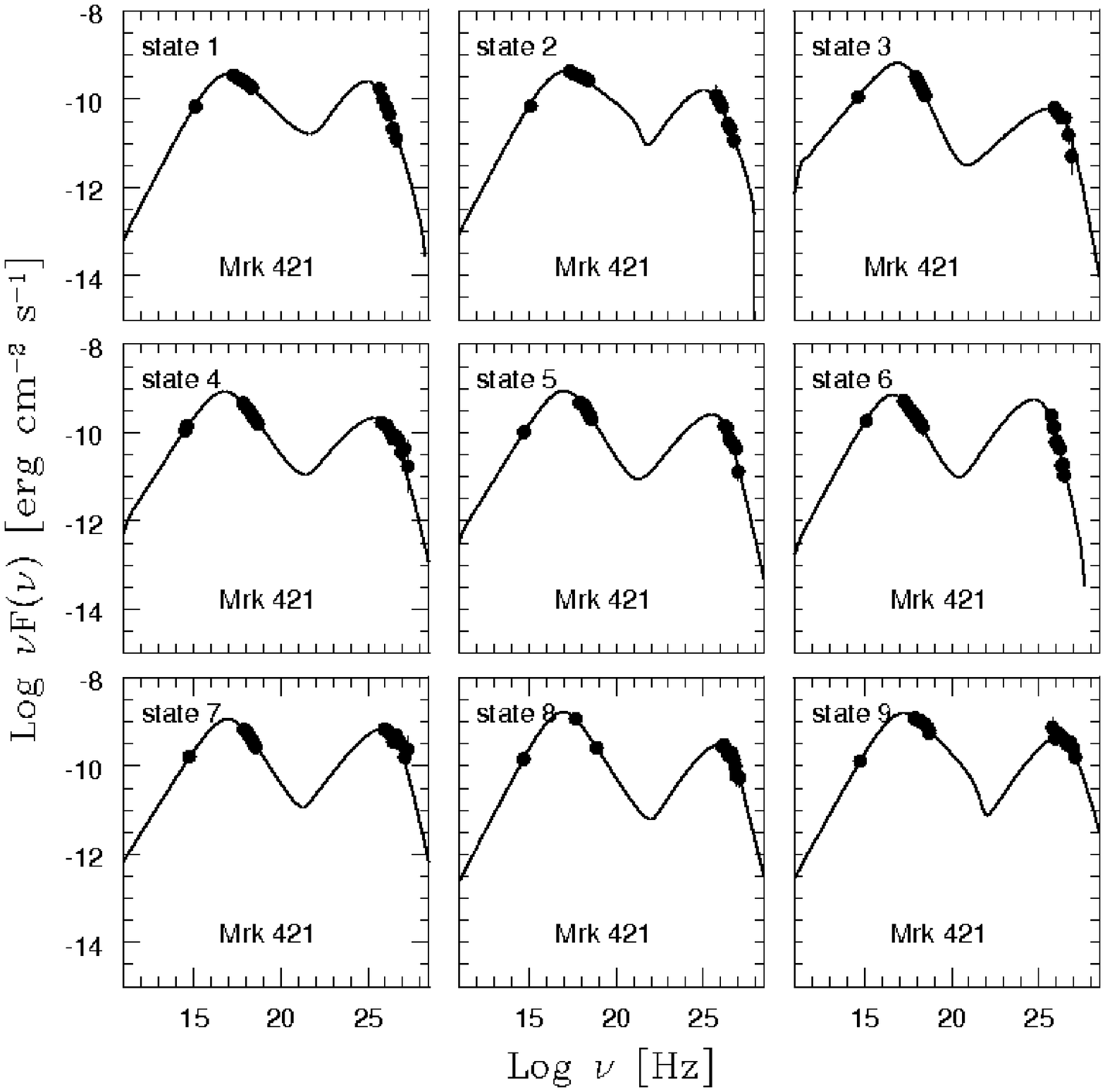}
\caption{Mkn\,421 in different emission states.
    }
  \includegraphics[width=5in,height=5in]{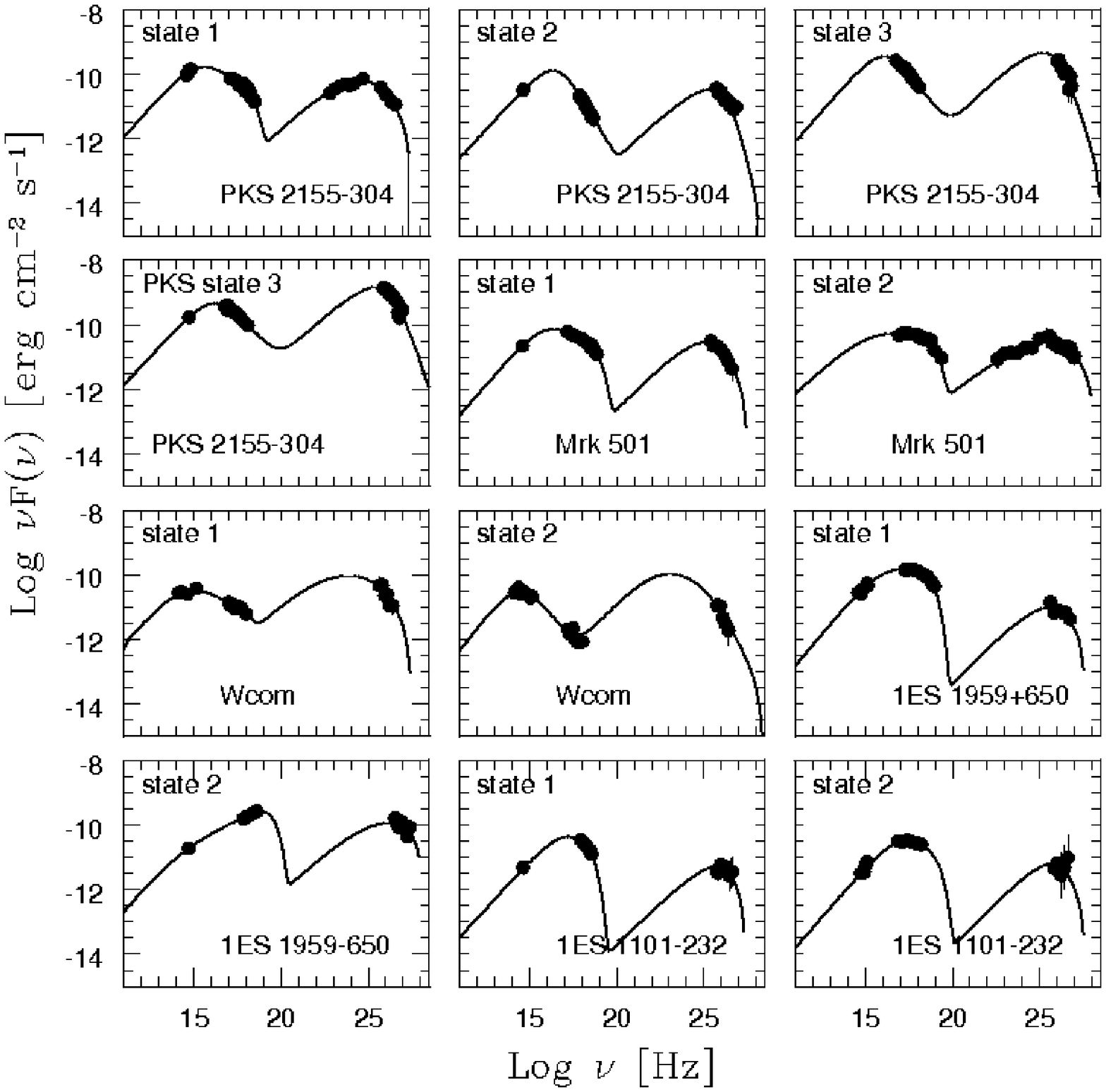}
  \caption{Other BL Lac objects (listed in Table 1) in different emission states.
    }
  \label{wide_fig}
 \end{figure*}

\section{Results}

\begin{figure}[!t]
  \vspace{-0.5cm}
  \centering
   \includegraphics[width=2.5in]{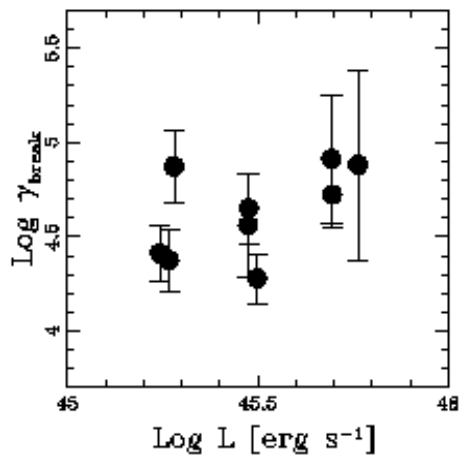}
\includegraphics[width=2.5in]{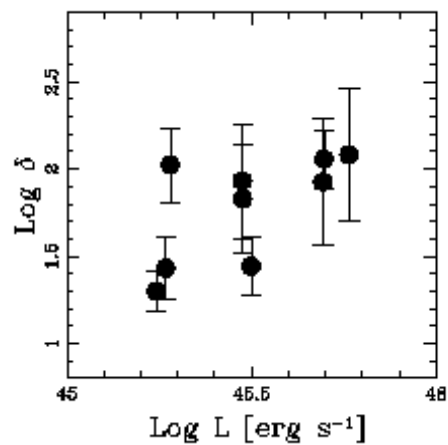}
\includegraphics[width=2.5in]{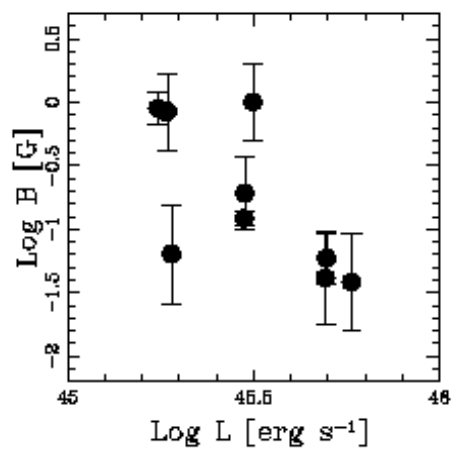}

  \caption{The SSC fit parameters $\gamma_{break}$, $\delta$ and $B$ versus source luminosity in Mkn\,421}
  \label{simp_fig}
 \end{figure}

Our results suggest that in Mkn\,421, $B$ decreases with source activity whereas $\gamma_{\rm break}$ and $\delta$ increase (Fig.\,3 top). This can be interpreted in a framework in which the synchrotron power and peak frequency remain constant with varying source activity by decreasing $B$ and increasing the number of low-energy electrons. This mechanism results in an increased electron-photon scattering efficiency and hence in an increased Compton power. Other emission parameters appear uncorrelated with source activity. In Fig.\,3 (bottom), the $B$-$\gamma_{\rm break}$ anti-correlation results from a roughly constant synchrotron peak frequency. The $B$-$\delta$ correlation suggests that the Compton peak of Mkn\,421 is always in the Thomson limit. The $\delta$-$\gamma_{\rm break}$ correlation is an effect of the constant synchrotron and Compton frequencies of the radiation emitted by a plasma in bulk relativistic motion towards the observer.

The analysis of the remaining sources is in progress. However, it should be noted that simultaneous MWL campaigns for other blazars are fewer than for Mkn\,421. Hence we encourage more MWL campaigns on blazars at their different emission level.

\begin{figure}[!t]
  \vspace{5mm}
  \centering
  \includegraphics[width=2.5in]{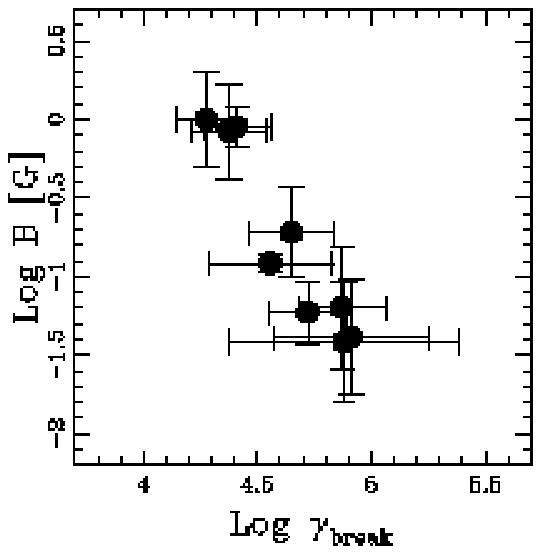}
\includegraphics[width=2.5in]{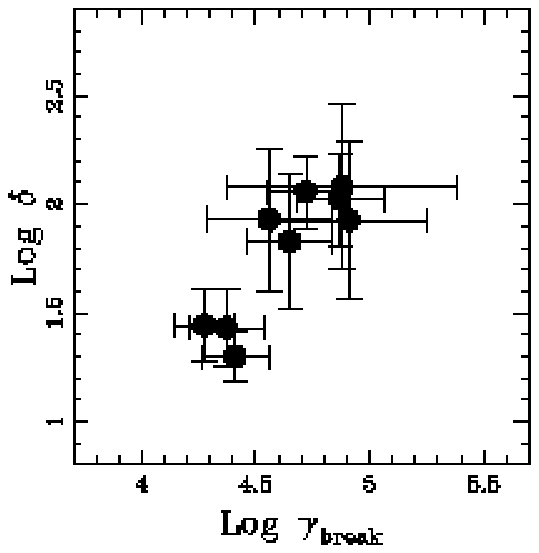}
\includegraphics[width=2.5in]{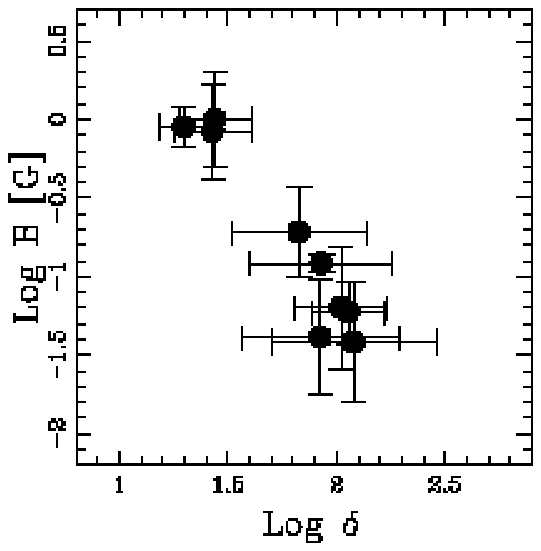}
  \caption{The correlation between the SSC fit parameters in Mkn\,421}
  \label{simp_fig}
 \end{figure}

To test the goodness of the fits we have applied the Kolmogorov-Smirnov (KS) test for normality of the residuals to all the SED fits presented in this paper. A standard application of the test shows that the residuals are not normally distributed in all cases, i.e. for most samples the KS test fails at 5\% significance. This could happen because of two reasons. (i) The one zone SSC model, which contains two distinct physical processes (synchrotron emission and its Compton upscattered counterpart) may only be an approximation to the real process: additional subtle effects
may also enter the modeling of blazar emission. (ii) In our blazar data sets the uncertainties associated with $\gamma$ data are much larger than those associated with the optical and X-ray data.
Both these observations suggest that the test could give different results if applied separately at low- and high-energy data. We found that even though the KS test fails on the whole data sets, it instead confirms normality of the residuals at 5\% confidence level when applied piecewise to low and high energy data.

With better and more complete VHE data, it would be possible to reduce the uncertainties in the parameters, to obtain a higher overall significance in the fit and, therefore, a clearer picture of the changes and properties and environment of the emitting electrons as a function of source activity.

\clearpage


\begin{thebibliography}{}

\bibitem{ref1} Blandford, R.D. \& Znajek, R.L., 1997, MNRAS, 179, 433

\bibitem{ref2} Maraschi, L. Ghisellini, G. \& Celotti, A., 1992, ApJ, 397, 5

\bibitem{ref3} Tavecchio, F., Maraschi, L., \& Ghisellini, G., 1998, ApJ, 509, 608

\bibitem{ref4} Press, W.H., et al. 1992, Numerical Recipes (Cambridge: Cambridge University Press)

\bibitem{ref5} Anderhub, H., et al., 2009, ApJ, 705, 1624

\bibitem{ref6} Abdo, A. A., et al., 2011, ApJ, 727, 129

\bibitem{ref7}Aharonian, F., et al., 2009, ApJL, 696, 150

\bibitem{ref8}Aharonian, F., et al., 2009, A \& A, 502, 749

\bibitem{ref9}Acciari, V. A., et al., 2009, ApJ, 707, 612

\bibitem{ref10}Acciari, V. A., et al., 2008, ApJL., 648, 73

\bibitem{ref11}Tagliaferri, G., et al., 2008, ApJ, 679, 1029

\bibitem{ref12}Krawczynski, H., et al., 2004, ApJ, 601, 151

\bibitem{ref13}Aharonian, F., et al., 2007, A\&A, 470, 475

\bibitem{ref14}Mankuzhiyil, N., Ansoldi, S., Persic, M., Tavecchio, F., 2011, ApJ, 733, 14











\end{thebibliography}
\end{document}